# Electrostatic potential mapping at ferroelectric domain walls by low-temperature photoemission electron microscopy

J. Schaab,[1] K. Shapovalov,[2] P. Schoenherr,[1] J. Hackl,[3] M. I. Khan,[3] M. Hentschel,[4] Z. Yan,[5,6] E. Bourret,[6] C. M. Schneider,[3,7] S. Nemsák,[3,8] M. Stengel,[2,9] A. Cano,[10] and D. Meier[1,11,a]

[1]Department of Materials, ETH Zurich, Vladimir-Prelog-Weg 4, 8093 Zurich, Switzerland

[2]Institut de Ciència de Materials de Barcelona (ICMAB-CSIC), Campus UAB, 08193 Bellaterra, Spain

[3]Forschungszentrum Jülich Peter Grünberg Institute (PGI-6), Leo-Brandt-Strasse, 52425 Jülich, Germany

[4]4th Physics Institute and Research Center SCoPE, University of Stuttgart, Pfaffenwaldring 57, 70569 Stuttgart, Germany

[5]Department of Physics, ETH Zurich, Otto-Stern-Weg 1, 8093 Zurich, Switzerland

[6]Materials Science Division, Lawrence Berkeley National Laboratory, Berkeley, California 94720, United States

[7]Physics Department, UC Davis, Davis, CA 95616, United States

[8]Advanced Light Source, Lawrence Berkeley National Laboratory, 1 Cyclotron Rd, Berkeley, CA, 94720, USA

[9]ICREA – Instució Catalana de Recerca i Estudis Avançats, 08010 Barcelona, Spain

[10]Institut Néel, CNRS & Univ. Grenoble Alpes, 38042 Grenoble, France

[11]Department of Materials Science and Engineering, Norwegian University of Science and Technology, NTNU, 7043 Trondheim, Norway

[a]Author to whom correspondence should be addressed. Electronic mail: dennis.meier@ntnu.no

Low-temperature X-ray photoemission electron microscopy (X-PEEM) is used to measure the electric potential at domain walls in improper ferroelectric $Er_{0.99}Ca_{0.01}MnO_3$. By combining X-PEEM with scanning probe microscopy and theory, we develop a model that relates the detected X-PEEM contrast to the emergence of uncompensated bound charges, explaining the image formation based on intrinsic electronic domain-wall properties. In contrast to previously applied low-temperature electrostatic force microscopy (EFM), X-PEEM readily distinguishes between positive and negative bound charges at domain walls. Our study introduces an X-PEEM based approach for low-temperature electrostatic potential mapping, facilitating nanoscale spatial resolution and data acquisition times in the order of 0.1–1 sec.

Fueled by the idea of domain-wall-based nanoelectronics[1-6], tremendous progress has been made over the last decade in measuring the nanoscale physical properties at ferroelectric domain walls[7,8]. Particularly interesting are so-called charged domain walls in ferroelectrics which, depending on their charge state (positive or negative) and the applied electric field, can either be insulating or conducting compared to the surrounding bulk[16]. This foreshadows the possibility to create, e.g., atomic scale wires, switches, gates and transistors[4,9,10]. The progress in the field is enabled by the continuous advancement of established high-resolution imaging techniques, as well as the development of new microscopy approaches. Modern scanning transmission electron microscopes, for example, allow to measure the atomic and chemical structure at domain walls with atomic scale precision[9]. Scanning probe and scanning electron microscopy (SPM and SEM) have become mainstream techniques for accessing domain-wall related transport phenomena[11-18], electrostatics[19,20] and magnetism[21]. More recently, X-ray photoemission electron microscopy (X-PEEM) and low energy electron microscopy (LEEM) have been employed to study emergent domain wall physics with nanoscale spatial resolution[22-24].

Despite the growing diversity of microscopy methods, the more fundamental investigation of domain walls at low temperatures remains surprisingly difficult. Pioneering studies mostly rely on low-temperature SPM, which enabled the observation of conductance at domain walls in Pb(Zr,Ti)$O_3$[25] and $Nd_2Ir_2O_7$[26], as well as domain-wall magnetism[21] and uncompensated bound charges[20] in $Er_{0.99}Ca_{0.01}MnO_3$. However, in order to go beyond these SPM-based breakthroughs, expand the currently accessible parameter space, and gain additional insight into the domain-wall nanoscale physics at cryogenic temperature, the introduction of additional low-temperature microscopy approaches is highly desirable.

Here, we demonstrate low-temperature X-PEEM as a powerful imaging and characterization tool for ferroelectric domain walls, providing so far inaccessible information about their electrostatic properties. Combining X-PEEM and SPM, we show that distinct X-PEEM contrasts arise at low temperature at positively (head-to-head) and negatively (tail-to-tail) charged domain walls in the improper ferroelectric $Er_{0.99}Ca_{0.01}MnO_3$[19,27], which we attribute to uncompensated polarization charges. In advantage to previous EFM studies[20], the X-PEEM experiment can distinguish between emergent positive and negative potentials and, hence, unambiguously identify head-to-head and tail-to-tail walls without the need of supplementary piezoresponse force microscopy (PFM) measurements. Furthermore, data acquisition times are in the order of 0.1–1 s, i.e., more than 100 times faster than the line-by-line image formation in EFM.

For our low-temperature X-PEEM experiments, high-quality ErMnO$_3$ and Er$_{0.99}$Ca$_{0.01}$MnO$_3$ single crystals are grown by the pressurized floating-zone method[28], oriented by Laue diffraction, and cut into platelets with the spontaneous polarization (P∥[001]) lying in the surface plane. The samples have a lateral dimension of about 4 mm and a thickness of 1 mm. To minimize topographical artifacts, the surface is chemo-mechanical polished, leading to a roughness of about 2 nm (RMS).

Figures 1(a) and 1(b) show PFM images (in-plane contrast) of the ferroelectric domains in ErMnO$_3$ and Er$_{0.99}$Ca$_{0.01}$MnO$_3$, respectively. Both samples exhibit the six-fold distribution of +P (dark) and −P (bright) domains characteristic for hexagonal manganites. Room-temperature X-PEEM images, gained on the same samples, are presented in Figure 1(c) and 1(d). The data in Figure 1(c) is obtained at a photon energy of 641.5 eV (Mn L$_3$ edge), showing several pronounced bright lines. These lines correspond to conducting tail-to-tail walls in ErMnO$_3$. These walls become visible in X-PEEM due to photo-induced charging as explained in detail in ref 22.

The photo-induced charging is less pronounced at conducting tail-to-tail walls than for the insulating bulk and head-to-head walls. As a consequence, photo-excited electrons from the tail-to-tail walls have a markedly higher kinetic energy, which can be visualized by adjusting the microscope´s energy filter[22]. In contrast to ErMnO$_3$, a homogenous energy distribution is observed for the photo-excited electrons in Er$_{0.99}$Ca$_{0.01}$MnO$_3$ (Figure 1(d), gained at 643.2 eV). This observation is consistent with the significantly higher bulk conductivity of Ca-doped ErMnO$_3$[19,27], which forbids the photo-induced surface charging, thus suppressing the formation of charging-induced domain-wall contrast at room-temperature.

At low-temperature, a distinctly different behavior is observed in Er$_{0.99}$Ca$_{0.01}$MnO$_3$ as displayed in Figure 2. After cooling to 160 K, both bright and dark lines are resolved in X-PEEM (Figure 2(a)). The image in Figure 2(a) is recorded after optimizing the X-ray energy to 643.1 eV to compensate for shifts in the Mn L$_{2,3}$ edge so that maximum contrast is achieved. Pt-markers designed using electron-beam lithography allow us to study the same sample position in subsequent cAFM and PFM scans presented in Figure 2(b) and 2(c). The comparison with the cAFM and PFM data identifies the bright and dark lines in the low-temperature X-PEEM image as tail-to-tail and head-to-head domain walls, respectively. Thus – different from the charging-induced contrast observed in ErMnO$_3$ at room-temperature (Figure 1(c)) – both conducting and insulating domain walls are resolved in the low-temperature X-PEEM images gained on Er$_{0.99}$Ca$_{0.01}$MnO$_3$. This qualitative difference suggests that the low-temperature behavior is dominated by a mechanism other than photo-induced charging.

A detailed investigation of the photoelectron energy distribution at tail-to-tail and head-to-head walls ($T = 160$ K) is displayed in Figure 2(d) and 2(e). The contour-plot in Figure 2(d) shows the normalized energy-dependent electron yield measured along the cross section marked in Figure 2(a). The energy-dependent data is gained by detecting local intensity variations while changing the so-called start energy $U_{st}$ (the center energy of a band pass energy filter). This allows quantification of differences in the kinetic energy of photo-excited electrons[22]. Maxima of the electron distribution (light blue) are found to shift in energy at the position of the domain walls. At head-to-head walls, the electron distribution shifts to lower energies with respect to the bulk, whereas a shift to higher energies occurs at tail-to-tail walls. The latter is highlighted in Figure 2(e), revealing an energy shift of 2.3 ± 0.2 eV between head-to-head and tail-to-tail walls.

Next, we consider the temperature-dependent evolution of the domain-wall signals. Figure 3 shows kinetic-energy maps obtained while heating from 160 K to room temperature. Maps of the spatial energy distribution are generated by fitting the secondary electron distribution for each pixel using a Gaussian fit function. At 160 K, pronounced energy shifts are detected, which correlate with the position of the domain walls, i.e., reduced and increased kinetic energy near head-to-head and tail-to-tail domain walls, respectively. With increasing temperature, the energy contrast gradually weakens until at 181 K only the head-to-head walls (blue) are resolved in the X-PEEM images. This is exactly the opposite compared to X-PEEM maps gained on ErMnO$_3$ at room temperature (Figure 1(c) and ref. 22), where only tail-to-tail walls are observed. This difference corroborates our assumption that different mechanisms are responsible for the X-PEEM contrast at low temperature; in particular, it allows for excluding photo-induced charging as the dominant effect.

To understand the detected X-PEEM contrasts at low temperature, we consider the intrinsic properties of the system. Recent low-temperature EFM studies revealed the emergence of uncompensated bound charges at head-to-head and tail-to-tail walls in Er$_{0.99}$Ca$_{0.01}$MnO$_3$ upon cooling[20], caused by the disruption of the internal screening mechanism at low temperatures. By decreasing the temperature, the spontaneous polarization of the domains increases due to the pyroelectric effect. This further increases the bound charge at both head-to-head and tail-to-tail domain walls, requiring an extra screening by mobile carriers of the bulk domains. As the effectiveness of the bulk screening is determined by mobile carriers, the screening mechanism can be disrupted due to a deficiency of mobile carriers below some characteristic temperature $T^*$ (note that the carrier density $n_h$ in the p-type semiconductor Er$_{0.99}$Ca$_{0.01}$MnO$_3$ decreases exponentially, i.e., $n_h \propto \exp[-E_F/kT]$, where $k$ is the Boltzmann constant and $E_F$ the distance between the Fermi energy and the top of the valence band)[20]. After cooling to $T < T^*$, electrostatic equilibrium is no longer reached within the timeframe of the experiment so that extra domain-wall charges remain uncompensated, leading to pronounced electrostatic fields.

In order to test whether such uncompensated domain-wall charges can explain the low-temperature X-PEEM data, we build a model that combines previous calculations of the formation[29] of domains and domain-wall bound charges[20] in hexagonal manganites, assuming the same bound charge screening behavior at head-to-head and tail-to-tail walls. For $T > T^*$, the material is in electrostatic equilibrium and we expect a mostly flat surface potential, which is consistent with the experimental result in Figure 1(d). At $T < T^*$, however, uncompensated charges appear at the domain walls with maximal surface density $\sigma = \pm 2p(T^* - T)$, where $p$ is the pyroelectric constant. The latter generates an electric potential $\phi(x, y)$ at the surface that shifts the kinetic energy of emitted electrons by $-e\phi(x, y)$. Figure 4 shows a Landau-theory-based simulation that illustrates the typical distribution of charged walls in hexagonal manganites[16] (Figure 4(a)) and the corresponding electric potential $\phi(x, y)$ for incomplete screening (Figure 4(b)) [see Supporting Note 1 for details of the calculations]. White (dark) shades correspond to a lower (higher) electric potential and, therefore, to higher (lower) kinetic energies of emitted electrons. Most importantly, the calculation reproduces the main experimental feature, that is, *opposite contrast* at head-to-head and tail-to-tail walls (Figure 3 and 4), suggesting that uncompensated bound charges do indeed play a key role for the X-PEEM image formation at low temperature.

To show the importance of uncompensated bound charges we make a numerical comparison of the simulated domain-wall contrast with the recorded X-PEEM data, focusing on regions with parallel domain walls. Figure 5(a) presents the relative energy shift of electrons emitted from tail-to-tail and head-to-head walls at different temperatures, corresponding to the cross-section marked in the inset to Figure 5(b). Figure 5(a) shows a pronounced asymmetry between head-to-head and tail-to-tail walls concerning the sign and magnitude of the detected energy shift, as well as its thermal evolution. The observed saw-tooth profile resembles the classic evolution of the electric potential between charged capacitor plates and its magnitude steadily increases as the temperature decreases. In Figure 5(b), we compare the thermal evolution of the minima (head-to-head) and maxima (tail-to-tail) in Figure 5(a) with the temperature scaling behavior expected from the model of a plate capacitor, assuming the same maximum density of domain-wall surface charge $\sigma$ as before. Dashed lines in Figure 5(b) are fitted to the X-PEEM data with $-e\Delta\phi = \pm epL(T^* - T)/2\epsilon$, where $L$ is the distance between the domain walls and $\epsilon$ is the dielectric constant (we note that the data point obtained for the head-to-head wall at 160 K is not considered for the fit as the line plot in Figure 5(a) indicates substantial interference with the signal associated with the tail-to-tail wall). Using $L = 800$ nm and $\epsilon = 13\epsilon_0$,[30] we extract the hypothetical pyroelectric constant of the material, $p = 2.5$ nC/(cm$^2$ K). This value is in remarkable agreement with literature data for the pyroelectric constants in other manganites [$p(HoMnO_3)$ = 0.6 nC/(cm$^2$ K)[31], $p(YMnO_3)$ = 3 nC/(cm$^2$ K)[32]], which corroborates our bound-charge model. Furthermore, the observed persistence of contrast at the head-to-head domain walls towards higher temperatures than for the tail-to-tail walls, and their overall offset of the energy shift (≈ -0.4 eV), are in qualitative agreement with the variations of the surface electric potential due to the hole depletion layer surrounding them[9,20] (see Supporting Note 2 for details). Interestingly, the extracted $T^*$ = 175 K is higher than in previous EFM experiments ($T^*$ = 120 K[20]), suggesting that X-PEEM is the more sensitive technique for electrostatic potential mapping. This higher sensitivity can be partially attributed to the fact that, in contrast to the electric far fields probed in EFM (typical distance $d \approx 30$ nm), the much stronger electric field at the sample surface ($d \to 0$) is responsible for image formation in X-PEEM[22]. Differences in the environmental conditions are another important factor: While the EFM data in ref. 20 has been recorded in He gas (20 mbar), X-PEEM is performed under ultrahigh vacuum (6 × 10$^{-10}$ mbar), which reduces signal loss due to extrinsic screening by surface contamination/adsorbates. In addition, as our estimates show, surface charging and the time-dependent nature of screening disruption can contribute to the higher $T^*$, causing a shift in the order of a few Kelvin for realistic material parameters. In conclusion, the numerical comparison of simulated and recorded X-PEEM contrasts (Figure 5) demonstrates that the low temperature potential maps gained on $Er_{0.99}Ca_{0.01}MnO_3$ are dominated by the emergence of uncompensated domain-wall charges.

In summary, we have applied low-temperature X-PEEM to detect uncompensated bound-charges at ferroelectric head-to-head and tail-to-tail domain walls in $Er_{0.99}Ca_{0.01}MnO_3$. Obtained X-PEEM maps and their temperature-dependent evolution is explained based on the charge-relaxation time and the freezing of mobile carriers at low temperature. In advantage to previously applied low-temperature EFM experiments, X-PEEM readily distinguishes between positive and negative bound charges, offering high sensitivity and substantially shorter data acquisition times. This work thus introduces X-PEEM as a valuable tool for the investigation of charged domain walls and for electrostatic potential mapping in general, expanding the limited set of imaging experiments that allows for electronic property characterization at low temperature and with nanoscale spatial resolution.

**ACKNOWLEDGMENTS**

We thank HZB for the allocation of synchrotron beam time and we gratefully acknowledge financial support by HZB. J.S. and D.M. acknowledge funding from the ETH Zurich and the SNF (proposal no. 200021_149192). D.M. thanks NTNU for support through the Onsager Fellowship Programme and NTNU Stjerneprogrammet. M.S. and K.S. acknowledge the support of the European Research Council under the European Union's Horizon 2020 research and innovation program (Grant Agreement No. 724529), of Ministerio de Economia, Industria y Competitividad through Grants No. MAT2016-77100-C2-2-P and No. SEV-2015-0496 and of the Generalitat de Catalunya (Grant No. 2017SGR 1506). M.H. acknowledges funding from ERC (Complexplas), DFG, BW Stiftung, and MWK BW (ZAQuant, IQST). Z.Y and E.B. were supported by the U.S. Department of Energy, Office of Science, Basic Energy Sciences, Materials Sciences and Engineering Divisionunder Contract No. DE-AC02-05-CH11231 within the Quantum Materials program-KC2202.

**REFERENCES**

1. E. K. H. Salje, ChemPhysChem **11**, 940 (2010).
2. E. K. H. Salje and H. Zhang, Phase Transitions **82**, 452 (2009).
3. G. Catalan, J. Seidel, R. Ramesh, and J. F. Scott, Rev. Mod. Phys. **84**, 119 (2012).
4. J. Seidel, L. W. Martin, Q. He, Q. Zhan, Y.-H. Chu, A. Rother, M. E. Hawkridge, P. Maksymovych, P. Yu, M. Gajek, et al., Nat. Mater. **8**, 229 (2009).
5. J. R. Whyte, R. G. P. McQuaid, P. Sharma, C. Canalias, J. F. Scott, A. Gruverman, and J. M. Gregg, Adv. Mater. **26**, 293 (2014).
6. J. R. Whyte and J. M. Gregg, Nature Commun. **6**, 7361 (2015).
7. D. Meier, J. Phys. Cond. Matter **27**, 463003 (2015).
8. P. S. Bednyakov, B. I. Sturman, T. Sluka, A. K. Tagantsev, and P. V. Yudin, npj Comp. Mater. **4**, 65 (2018).
9. J. A. Mundy, J. Schaab, Y. Kumagai, A. Cano, M. Stengel, I. P. Krug, D. M. Gottlob, H. Doğanay, M. E. Holtz, R. Held, et al., Nat. Mater. **16**, 622 (2017).
10. J. Schaab, S.H. Skjærvø, S. Krohns, X. Dai, M. E. Holtz, A. Cano, M. Lilienblum, Z. Yan, E. Bourret, D. A. Muller, et al., Nat. Nano. **13**, 1028 (2018).
11. V. V. Aristov, L. S. Kokhanchik, and Y. I. Voronovskii, Phys. Status Solidi A **86**, 133 (1984).
12. J. Li,, H. X. Yang, H. F. Tian, C. Ma, S. Zhang, Y. G. Zhao, and J. Q. Li, Appl. Phys. Lett. **100**, 152903 (2012).
13. T. Sluka, A. K. Tagantsev, D. Damjanovic, M. Gureev, and N. Setter, Nat. Commun. **3**, 748 (2012).
14. J. Guyonnet, I. Gaponenko, S. Gariglio, and P. Paruch, Adv. Mater. **23**, 5377 (2011).
15. S. Farokhipoor and B. Noheda, Phys. Rev. Lett. **107**, 127601 (2011).
16. D. Meier, J. Seidel, A. Cano, K. Delaney, Y. Kumagai, M. Mostovoy, N. A. Spaldin, R. Ramesh, and M. Fiebig, Nat. Mater. **11**, 284 (2012).
17. P. Maksymovych, A. N. Morozovska, P. Yu, E. A. Eliseev, Y.-H. Chu, R. Ramesh, A. P. Baddorf, and S. V. Kalinin, Nano Lett. **12**, 209 (2012).

18. M. Schröder, A. Haußmann, A. Thiessen, E. Soergel, T. Woike, and L. M. Eng, Adv. Funct. Mater. **22**, 3936 (2012).
19. J. Schaab, A. Cano, M. Lilienblum, Z. Yan, E. Bourret, R. Ramesh, M. Fiebig, and D. Meier, Adv. Electron. Mater. **2**, 1500195 (2016)
20. P. Schoenherr, K. Shapovalov, J. Schaab, Z. Yan, E. D. Bourret, M. Hentschel, M. Stengel, M. Fiebig, A. Cano, and D. Meier, Nano Lett. **19**, 1659 (2019).
21. Y. Geng, N. Lee N, Y. J. Choi, S.-W. Cheong, and W. Wu, Nano Lett. **12**, 6055 (2012).
22. J. Schaab, I. P. Krug, F. Nickel, D. M. Gottlob, H. Doğanay, A. Cano, M. Hentschel, Z. Yan, E. Bourret, C. M. Schneider, et al., Appl. Phys. Lett. **104**, 232904 (2014).
23. A.-S. Pawlik, T. Kämpfe, A. Haußmann, T. Woike, U. Treske, M. Knupfer, B. Büchner, E. Soergel, R. Streubel, A. Koitzsch, et al., Nanoscale **9**, 10933 (2017).
24. Z. Zhao, N. Barrett, Q. Wu, D. Martinotti, L. Tortech, R. Haumont, M. Pellen, and E. K. H. Salje, Phys. Rev. Materials **3**, 043601 (2019).
25. I. Stolichnov, L. Feigl, L. J. McGilly, T. Sluka, X.-K. Wie, E. Colla, A. Crassous, K. Shapovalov, P. Yudin, A. K. Tagantsev, et al., Nano Lett. **15**, 8049–8055 (2015).
26. E. Y. Ma, Y.-T. Cui, K. Ueda, S. Tang, K. Chen, N. Tamura, P. M. Wu, J. Fujioka, Y. Tokura, and Z.-X. Shen, Science, **350**, 538 (2015).
27. E. Hassanpour, V. Wegmayr, J. Schaab, Z. Yan, E. Bourret, Th. Lottermoser, M. Fiebig, and D. Meier, New. J. Phys. **18**, 43015 (2016).
28. Z. Yan, D. Meier, J. Schaab, R. Ramesh, E. Samulon, and E. Bourret, J. Cryst. Growth **409**, 75 (2015).
29. M. Holtz, K. Shapovalov, J. A. Mundy, C. S. Chang, Z. Yan, E. Bourret, D. A. Muller, D. Meier, and A. Cano, Nano Lett. **17**, 5883 (2017).
30. M. Stengel, C. J. Fennie, and P. Ghosez, Phys. Rev. B **86**, 094112 (2012).
31. N. Hur, I. K. Jeong, M. F. Hundley, S. B. Kim, and S.-W. Cheong, Phys. Rev. B **79**, 134120 (2009).
32. S. Artyukhin, K. Delaney, N. A. Spaldin, M. Mostovoy, Nat. Mater. **13**, 42 (2014).

**FIGURES**

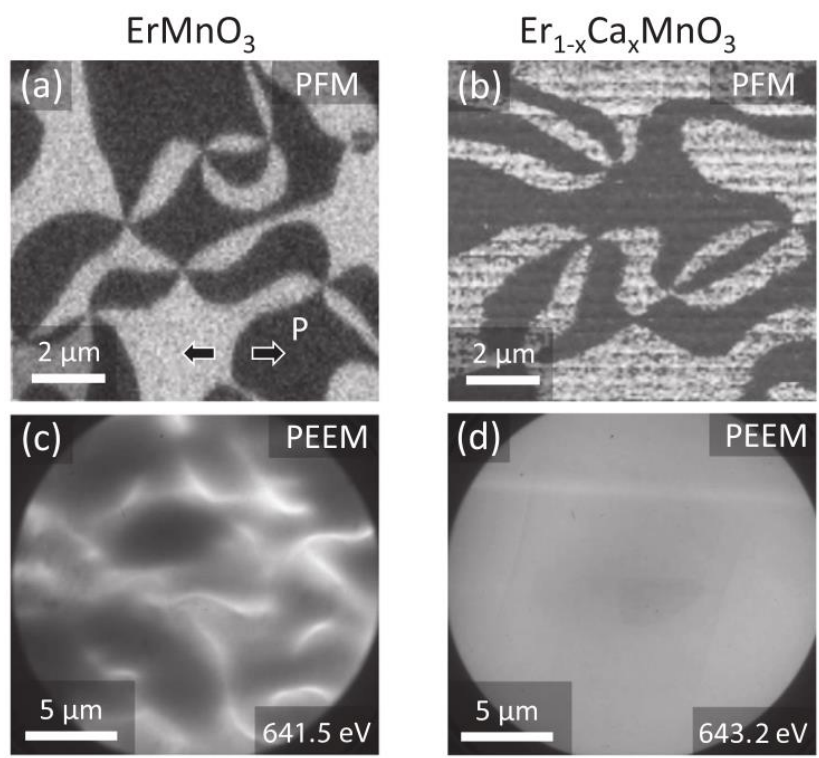

FIG. 1. (a),(b) PFM images (in-plane contrast) of ferroelectric domains in $ErMnO_3$ and $Er_{0.99}Ca_{0.01}MnO_3$. Arrows in (a) indicate the direction of the spontaneous polarization P. (c),(d) Room-temperature X-PEEM images at a photon energy of 641.5 eV (Mn $L_3$ edge). While bright charging-induced contrast is observed at tail-to-tail domain walls in $ErMnO_3$, a homogenous contrast level is obtained for $Er_{0.99}Ca_{0.01}MnO_3$.

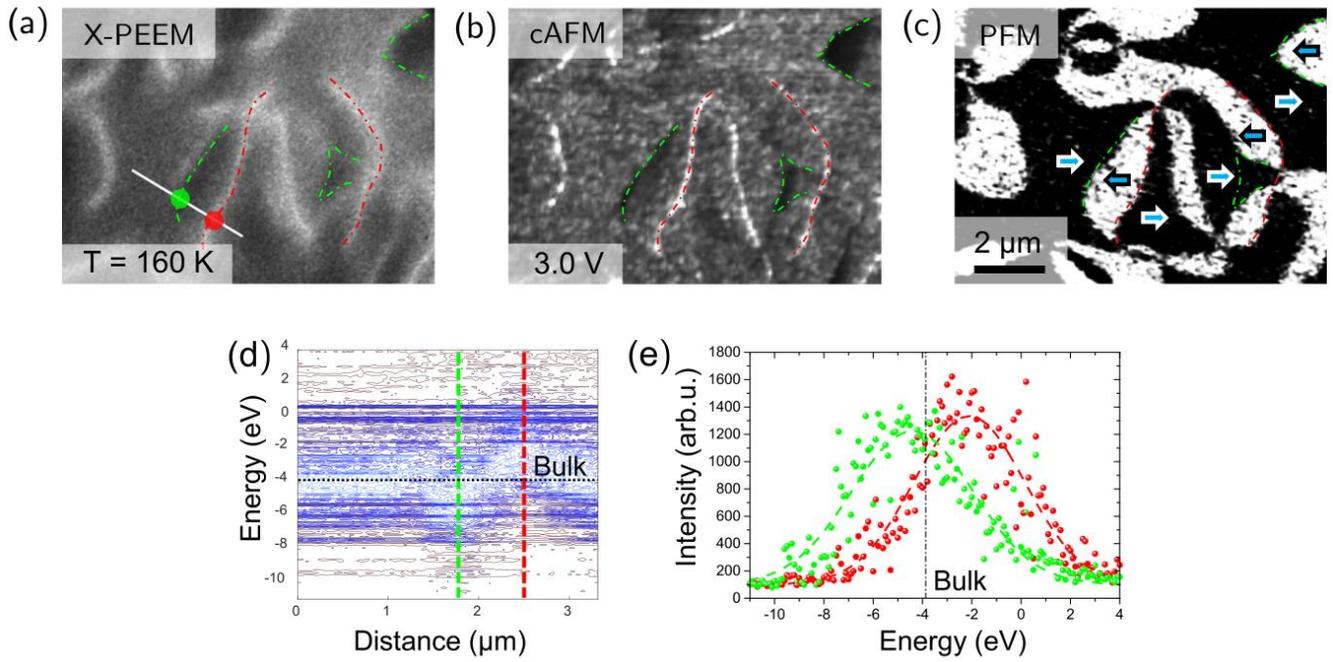

FIG. 2. (a), (b), and (c) compare X-PEEM, cAFM, and PFM images recorded in the same area on $Er_{0.99}Ca_{0.01}MnO_3$. The local conductance (cAFM) is measured at 3V bias applied to the back electrode; green and red dashed lines indicate the position of insulating head-to-head (dark) and conducting tail-to-tail (bright) walls. The X-PEEM data is taken at 160 K, whereas cAFM and PFM maps are recorded at room-temperature. (d) Normalized electron yield as function of electron energy at head-to-head (green dashed line) and tail-to-tail (red dashed line) walls measured along the cross section marked by the white dashed line in (a) (light blue = high energy, grey = low energy). (e) Electron energy distribution recorded at the wall positions marked in (a) and (d). Black dashed lines in (d) and (e) correspond to the average kinetic energy of electrons emitted from the bulk.

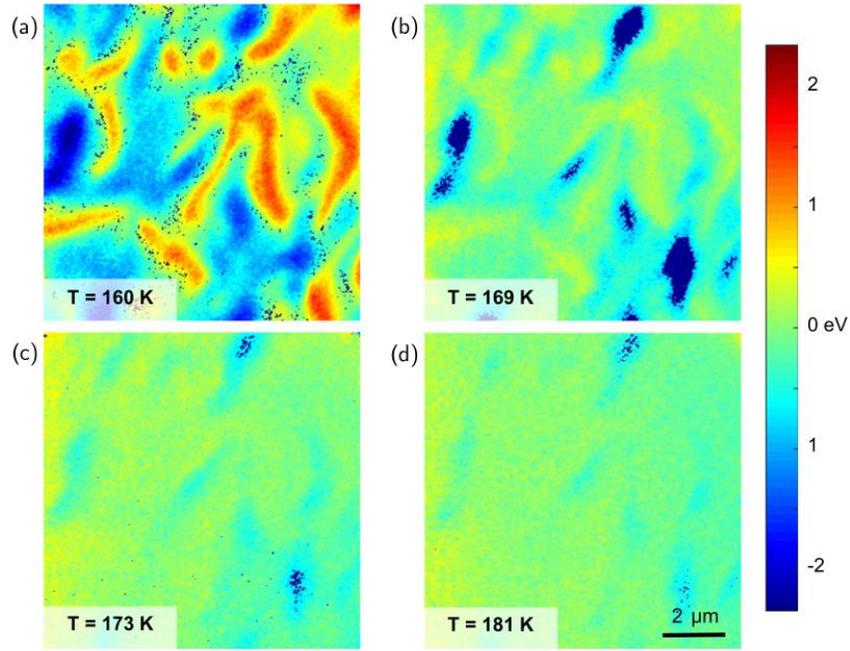

FIG. 3. (a) to (d) Temperature dependence of the spatial energy distribution of photo-excited electrons in $Er_{0.99}Ca_{0.01}MnO_3$. Colors reflect the maxima of the electron energy distribution as indicated on the right. The maximum energy is derived from an energy-dependent image series, evaluating the electron energy distribution pixel-by-pixel using a Gaussian fit function.

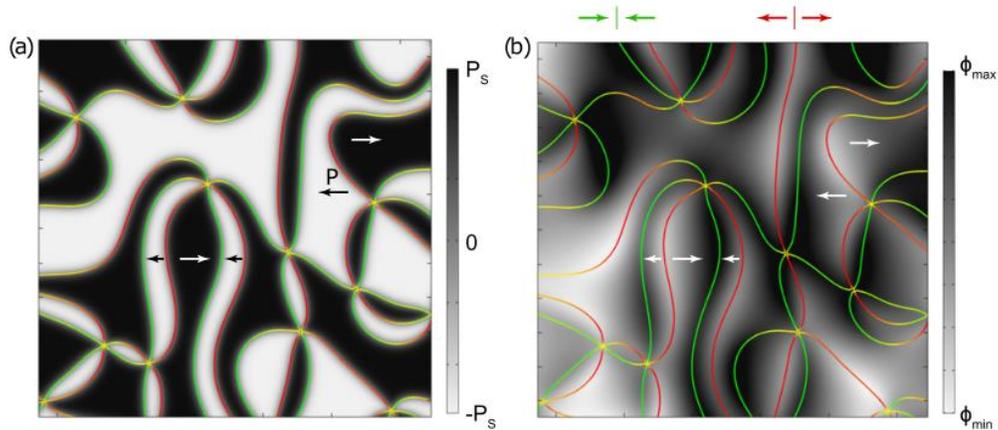

FIG. 4. (a) Calculated domain structure using the finite-elements method. (b) Corresponding map of surface electric potential generated by uncompensated bound charges at head-to-head (green) and tail-to-tail (red) domain walls. Potential maxima at head-to-head walls and minima at tail-to-tail walls correspond to negative and positive eV-shifts in X-PEEM images (see Figure 2 and 3).

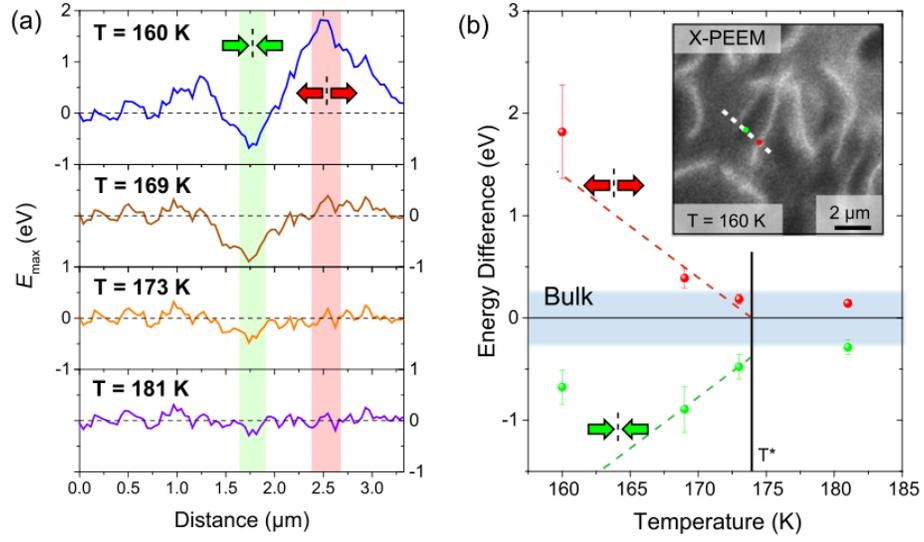

FIG. 5. (a) Energy profile of photoelectrons along the white-dashed line marked on the inset to (b). Positions of head-to-head and tail-to-tail walls are highlighted green and red, respectively. (b) Temperature dependence of the maxima/minima in (a) relative to the bulk. Blue color indicates the uncertainty in the average energy of electrons emitted from the bulk. Dashed lines are fits based on a plate capacitor model as explained in the main text (pyroelectric constant $p = 2.5$ nC cm$^{-2}$ K$^{-1}$).

**SUPPORTING INFORMATION**

**Supporting Note 1: Numerical modeling of the surface potential**

In order to simulate the X-PEEM response shown in the manuscript, we developed a numerical model implementing the finite-element method similar to Ref. 29. First, we minimize the Landau potential describing the space evolution of the trimerization order parameter $\mathbf{Q} = (Q\cos\Phi, Q\sin\Phi)$:

$$F = \frac{a}{2}Q^2 + \frac{b}{4}Q^4 + \frac{1}{6}(c + c'\cos 3\Phi)Q^6 + \frac{g}{2}((\nabla Q)^2 + Q^2(\nabla\Phi)^2), \quad (S.1)$$

using COMSOL Multyphisics. We start from an arbitrary smooth distribution of $\mathbf{Q}(x,y)$ with set characteristic length $L_c$ of variation of $\mathbf{Q}(x,y)$, and let the trimerization order parameter relax until a metastable domain structure is reached as shown in Figure 4(a) of the manuscript. The length $L_c$, translating into the characteristic domain size in the relaxed structure, was chosen $\sim 10 \times$ domain wall width to optimize the computation time while still maintaining the condition that the domain size is much larger than the domain wall width.

In our model, assuming the same screening behaviour at head-to-head and tail-to-tail walls, the electric potential is generated by unscreened bound charges at the domain walls, which change with temperature as $p(T - T^*)$ [$p$ is the pyroelectric constant]. Therefore, the electric potential scales linearly with temperature: $\phi(x,y) = \phi_0(x,y)(T^* - T)$. Figure 4(b) of the main text is obtained by feeding $P_x \propto Q^3 \cos(3\Phi)$ [Ref. 29] (where $x$ indicates the [001] direction) into the Poisson equation $\nabla(-\varepsilon\nabla\phi + \mathbf{P}) = 0$ and shows the distribution of the electric potential $\phi_0(x,y)$ without focusing on its absolute value.

**Supporting Note 2: Numerical modeling of the X-PEEM response of domain walls**

One can see in Figures 3 and 5 of the main text that head-to-head domain walls are detectable towards higher temperatures than tail-to-tail walls, with a $\sim 0.4$ eV offset. This asymmetry is in qualitative agreement with the fact that in $Er_{0.99}Ca_{0.01}MnO_3$ the bound charges at the head-to-head domain walls are screened by a hole depletion layer[9,20], producing a $\sim 1.6$ V peak of the electric potential at the domain wall. Ideally one would expect this peak to translate into $-1.6$ eV contrast in X-PEEM images, but as we show below, if the width of the depletion layer is smaller than the X-PEEM spacial resolution ($\sim 50$ nm), this peak will naturally flatten out in the images.

We develop a model where we simulate the X-PEEM response to the variations of the surface electric potential taking into account the X-PEEM spacial resolution. Focusing on flat regions of domain walls (Figure 5), we assume, similar to Ref. 20, that the electric potential is generated by two sources: domain wall bound charges as discussed in the main text and in the Supplementary Note 1, and the depletion layer surrounding the head-to-head domain wall. The first mechanism produces a saw-tooth potential between the domain walls (reaching maximum value of $\pm pL(T^* - T)/2\epsilon$ at the walls, where $L$ is the distance between the walls), while the second produces a 1.6 V peak localized within the depletion layer of width $2w = \sqrt{2\epsilon E_g/(e^2 N_A)}$, where $E_g$ is the band gap and $N_A$ is the density of acceptors – see Fig. S.1(a).

We fit the distribution of the kinetic energy of emitted electrons (Figure 2(e) of the main text) with the Gaussian function:

$$w_0(E, \phi) = \exp\left(-\frac{(E + e\phi - E_0)^2}{\sigma^2}\right), \qquad (S.2)$$

where $E_0$ is the median energy of electrons emitted from the bulk domain, and $\sigma$ is the dispersion. To model the non-zero spacial resolution of X-PEEM, we smear the distribution $w_0(E, \phi(x))$ over $\delta_{PEEM}$ by convoluting it with $\exp(-x^2/\delta_{PEEM}^2)$:

$$w(E, x) = \frac{1}{\sqrt{\pi \delta_{PEEM}^2}} \int dx_0 \, w_0(E, \phi(x_0)) \exp\left(-\frac{(x - x_0)^2}{\delta_{PEEM}^2}\right). \qquad (S.3)$$

Using the same parameters as in the manuscript ($L = 800$ nm, $\epsilon = 13\epsilon_0$, $p = 2.5$ nC/(cm$^2$K), $T^* = 174$ K), as well as $\sigma = 3$ eV, $E_g = 1.6$ eV, $N_A = 2 \times 10^{19}$ cm$^{-3}$, $2\delta_{PEEM} = 50$ nm, we reproduce the main features of X-PEEM shown in Figures 2(e) and 5 of the manuscript – cf. Fig. S.1(b,c).

**SUPPORTING FIGURE**

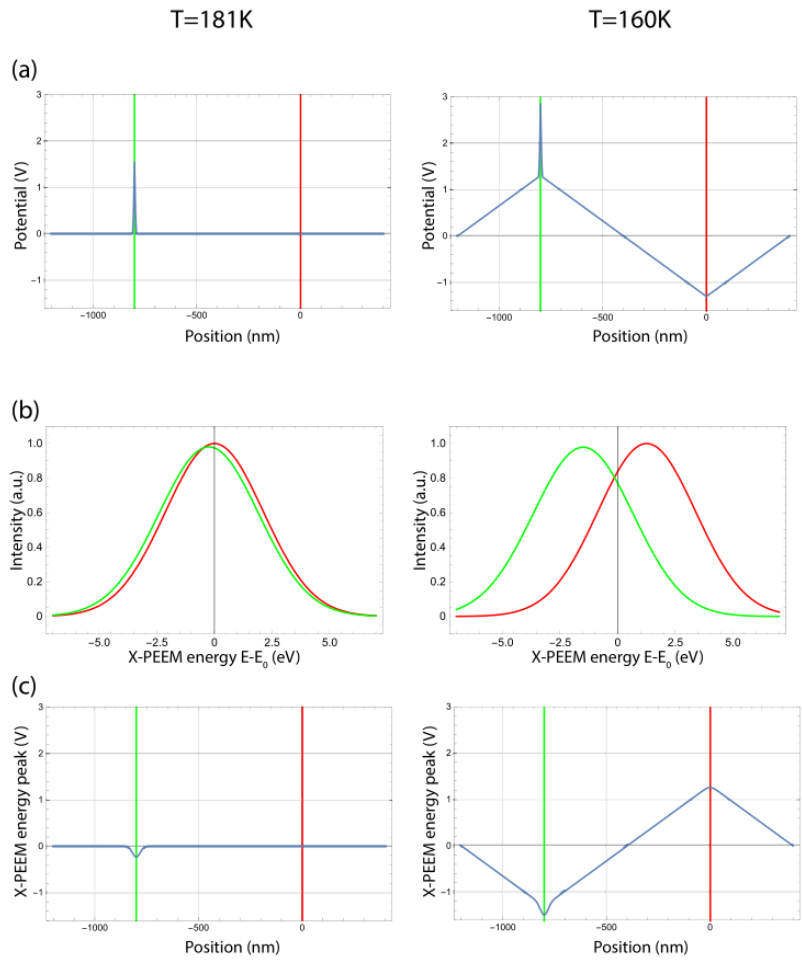

FIG. S.1. Simulation of X-PEEM response at two notable temperatures of the manuscript: 181 K ($T > T^*$) and at 160 K ($T < T^*$). (a) Modeled electric potential between head-to-head (green) and tail-to-tail (red) walls. (b) Modeled energy distribution of emitted electrons on head-to-head (green) and tail-to-tail (red) walls given by Eq. (S.3). (c) X-PEEM energy distribution maxima at different positions in the domain structure.